\documentclass{Rinton-P9x6}
\begin{document}
\title{Binary communication in entangled channels}
\author{Matteo G A Paris}
\address{Quantum Optics \& Information Group, 
INFM Unit\`a di Pavia, Italy \\ {\tt E-mail:~paris@unipv.it}, 
{\tt URL: www.qubit.it/\~{}paris}}
\maketitle
\abstracts{We analyze optical binary communication assisted by 
entanglement and show that: i) ideal entangled channels have 
smaller error probability than ideal single-mode coherent 
channels if the photon number of the channel is larger than one; 
ii) realistic entangled channels with heterodyne receivers have
smaller error probability than ideal single-mode coherent channels 
if the photon number of the channel is larger than a threshold 
of about five photons.} 
\section{Introduction}
In order to convey classical information to a receiver using 
quantum channels a transmitter prepares a quantum state drawn 
from a collection of known states. The receiver detects the 
information by measuring the channel, such to determine the state 
prepared. Since the given states are generally not orthogonal, 
then no measurement will allow the receiver to distinguish 
perfectly between them. The problem is therefore to construct 
a measurement optimized to distinguish between nonorthogonal 
quantum states, and to find realistic signals that minimize the 
error probability at fixed energy of the channel.
\par
In optical binary communication, say amplitude modulation keyed (AMK) 
\footnote{An equivalent analysis may be performed for phase-shift 
keyed signals.}, information is conveyed by two quantum states $\varrho_j=
|\psi_j\rangle\langle \psi_j|$, $j=1,2$ with $|\psi_1\rangle=
|\psi_0\rangle$ and $|\psi_2\rangle=D(\alpha)|\psi_0\rangle$ 
where $|\psi_0\rangle$ is a given reference state, usually taken as 
the vacuum. The amplitude $a$ may be taken as real without loss
of generality, and $D(\alpha)=\exp(\alpha a^\dag - \bar\alpha a)$ is
the displacement operator. If we consider equal {\em a priori} probabilities 
for the two signals, the optimal quantum measurement to discriminate the 
$|\psi\rangle$'s with minimum error probability \cite{hel} is the 
POVM $\{M_j\}_{j=1,2}$, $M_1+M_2=I$ corresponding to the so-called 
square-root measurement \cite{sqm}. We have $M_j= |\mu_j\rangle\langle
\mu_j |$ with $|\mu_j\rangle=\sum_k \mu_{kj}|k\rangle$, $\mu_{kj}=[\Psi 
(\Psi^\dag \Psi)^{-1/2}]_{kj}$. $[\Psi]_{ij}=\psi_{kj}$ is the matrix 
of the coefficients of the two signals $|\psi_j\rangle=\sum_k \psi_{kj} 
|k\rangle$ in a given basis $\{|k\rangle\}$. The error probability is given 
by $$P_e=\frac12\hbox{Tr}[M_1\varrho_2+M_2 \varrho_1]=\frac12 \left[1-
\sqrt{1-|\langle\psi_1 |\psi_2\rangle|^2}\right]\:.$$ For AMK binary 
communication we have $|\langle\psi_1|\psi_2\rangle|^2=\exp(-2N)$, where 
$N$ is the average number of photons in the channel {\em per use} 
$N=\frac12 \hbox{Tr}[a^\dag a (\varrho_1+\varrho_2)]=\frac12 |\alpha|^2$. 
In the following we will refer to this quantity as {\em the photon number of 
the channel}. \par 
In this communication we describe how binary communication can be improved
by using realistic sources of entanglement, either considering ideal or 
heterodyne receivers for state discrimination. The corresponding error 
probabilities are denoted by $Q_e$ and $R_e$ respectively. We find that
entanglement is convenient unless the photon number of the channel is very
small.  
\section{Binary communication in entangled channels}
Binary optical communication assisted by entanglement may be implemented using
as a reference state the so-called twin-beam (TWB) state of two modes of
radiation $|\psi_0\rangle=|\lambda\rangle\rangle=\sqrt{1- \lambda^2}\sum
\lambda^p\: |p\rangle|p\rangle$, $|\lambda|<1$.  TWBs are produced by
spontaneous downconversion in a nondegenerate optical parametric amplifier.
The TWB parameter is given by $\lambda=\tanh G$, $G$ being the effective gain
of the amplifier. The two states to be discriminated are now given by
$|\psi_1\rangle\rangle=|\lambda\rangle\rangle$ and
$|\psi_2\rangle\rangle=D(\alpha)|\lambda\rangle\rangle$, where we consider the
displacement performed on the beam $a$.
Since the average number of photons of TWB is given by $N_\lambda=
2\lambda^2/(1-\lambda^2)$, the photon number  of TWB channels is given by
$N=N_\lambda+\frac12 |\alpha|^2$. The error probability for the ideal
discrimination of the states $|\psi_1\rangle\rangle$ and
$|\psi_2\rangle\rangle$ reads as follows  $Q_e=\frac12
\left[1-\sqrt{1-\exp[-2N (1-\beta)(1+\beta N)]}\right]$, where
$\beta=N_\lambda/N$ is the fraction of the photon number of the channel used
to establish entanglement between the two modes.  We have that $Q_e<P_e$ for
$N>(1-\beta)^{-1}$ {\em i.e.} entanglement is always convenient if the photon
number of the channel is larger than one. The optimal entanglement fraction is
given by $\beta_{opt}=(N-1)/(2N)$, corresponding to an error probability given
by $Q_e=P_e$ if $N<1$ and 
$$Q_e=\frac12 \left[1-\sqrt{1-\exp[-\frac12 (1+N)^2]}\right]\qquad N\geq 1
\:.$$ \par
In assessing the usefulness of entanglement in binary communication
it should be taken into account that for the single-mode coherent 
channels the performances of the ideal detector can be in principle 
achieved by means of the Dolinar's receiver \cite{dol}. On the other 
hand, it has not been so far suggested how to achieve ideal performances 
for TWB entangled channels. Therefore, a question arises whether or not 
entanglement could be practically used to improve binary communication.
\par
\begin{figure}[h]
\psfig{file=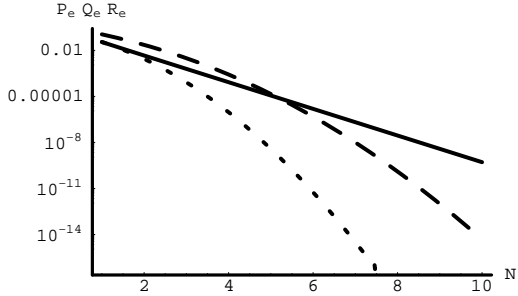,width=7cm}
\caption{Logarithmic plot of the error probabilities as a function of the 
photon number of the channel. The solid line is the error probability $P_e$ 
for a single-mode coherent channel, dotted line is $Q_e$ for ideal 
entangled channels, and dashed line for $R_e$ of heterodyne entangled 
channels. $R_e < P_e$ for $N$ larger than the threshold $N\simeq 5.2$.
\label{f:probs}} \end{figure}

In order to answer to this question we consider a receiver measuring 
the real and the imaginary part of the complex operator $Z=a+b^\dag$, 
$a$ and $b$ being the two modes of the TWB. The measurement of 
$\hbox{Re}[Z]$ and $\hbox{Im[Z]}$ can be experimentally implemented, 
and corresponds to multiport homodyne detection if the two involved 
modes have the same frequencies \cite{tri}, or to heterodyne 
detection otherwise \cite{het}. The outcome of each $Z$ measurement 
is a complex number $z$, and the POVM of the measurement is given by
$\Pi_z=|z\rangle\rangle\langle\langle z| $, with $|z\rangle\rangle =
D_j(z)\sum_n |n\rangle|n\rangle$ and either $j=a$ or $j=b$. 
As inference rule we adopt the condition "$\hbox{Re}[z]>\Lambda \rightarrow \: 
|\psi_2\rangle\rangle$", where $\Lambda$ is a threshold value, to be
determined such to minimize the probability of error $R_e$. Since the
heterodyne distribution, conditioned to a displacement $\alpha$, 
is given by $p(z|\alpha)=|\langle\langle z|D(\alpha)|\lambda\rangle
\rangle|^2= 1/\Delta_\lambda \exp\{-|\alpha-z|^2/
\Delta^2_\lambda\}$ with $\Delta^2_\lambda = (1-\lambda)/(1+\lambda)=
(\sqrt{N_\lambda+2}-\sqrt{N_\lambda})/(\sqrt{N_\lambda+2}+\sqrt{N_\lambda})$, 
we have $R_{12}=\int^{\hbox{Re}[z]>\Lambda}_0d^2 z \: p(z|\alpha)$ and 
$R_{21}=\int_{\hbox{Re}[z]>\Lambda}^\infty d^2z \:p(z|0)$ 
such that $$R_e=\frac12 (R_{12}+R_{21})=1- \frac12 \left\{\hbox{Erf}\left[
\frac{\Lambda}{\Delta_\lambda}\right]+ \hbox{Erf}\left[\frac{\alpha-
\Lambda}{\Delta_\lambda}\right]\right\}\:,$$ where $\hbox{Erf}[x]=
\frac2{\sqrt{\pi}}\int_0^x dt e^{-t^2}$ denotes the error function.  
$R_e$ is minimized by the choice $\Lambda=\alpha/2$, and therefore the minimum
probability of error in a realistic entangled channel with heterodyne receiver
is given by $R_e=\frac12 [1-\hbox{Erf}(\frac12 \alpha/\Delta_\lambda)]$.
Error probability $R_e$ can be further minimized by tuning the entanglement
fraction $\beta$. Substituting in $R_e$ the expression of 
$\alpha = \sqrt{2 N (1-\beta)}$ and $\Delta_\lambda$ we obtain 
$$R_e=\frac12\left\{ 1-\hbox{Erf}\left[\frac12
\sqrt{\frac{2N(1-\beta)(\sqrt{\beta N+2}-\sqrt{\beta N})}{(\sqrt{\beta
N+2}+\sqrt{\beta N})}} \right]\right\}\:.$$
The optimal entanglement fraction is given by 
$\beta_{opt}=\frac12 N/(1+N)$ which maximizes the argument of Erf, and thus 
minimizes $R_e$. 
We do not report the resulting error probability, whose expression is rather
cumbersome. Rather, in Fig. \ref{f:probs} we report the error probabilities as
a function of the photon number of the channel. We have that $R_e < P_e$ for
$N$ larger than the threshold value $N\simeq 5.2$. As it is
apparent from the plot, although TWB entangled channels with heterodyne
receiver do not approach the ideal performances, they definitely show smaller
error probability than single-mode coherent channels 
\par Asymptotically, for large photon number of the channel, we have 
$P_e\simeq \frac14 e^{-2N}$, $Q_e\simeq \frac14 e^{-\frac12 (1+N)^2}$ 
and $R_e\simeq (\sqrt{2\pi}N)^{-1} e^{-\frac18 N^2}$.
\section{Conclusions}
We have analyzed binary communication in TWB-based entangled 
channels, and compared their performances with ideal single-mode 
coherent channels. We have found that ideal entangled channels  
show smaller error probability than single-mode ones if the photon 
number of the channel is larger than one, whereas realistic entangled 
channels, {\em i.e.} channels equipped with heterodyne receivers, show 
smaller error probability than single-mode ones if the photon number 
of the channel is larger than a threshold of about five photons. 
We conclude the TWB binary communication represents a realistic 
alternative to single-mode coherent channels.
\section*{Acknowledgments}
This work has been sponsored by the INFM through the project PRA-2002-CLON 
and by EEC through the project IST-2000-29681 (ATESIT).

\end{document}